\documentclass[sigconf]{acmart}

\usepackage{multirow}
\usepackage{subcaption}
\usepackage{booktabs}
\usepackage{enumerate}
\usepackage{makecell}
\usepackage{tikz}
\usepackage{enumitem}
\usepackage{graphicx}
\usepackage{bm}
\usepackage{balance}

\AtBeginDocument{%
  }

\copyrightyear{2025}
\acmYear{2025}
\setcopyright{acmlicensed}
\acmConference[SIGIR '25] {Proceedings of the 48th International ACM SIGIR Conference on Research and Development in Information Retrieval}{ July 13--18, 2025}{Padua, Italy.}
\acmBooktitle{Proceedings of the 48th International ACM SIGIR Conference on Research and Development in Information Retrieval (SIGIR '25), July 13--18, 2025, Padua, Italy}
\acmISBN{979-8-4007-1592-1/25/07}
\acmDOI{10.1145/3726302.3730111}
\begin{document}

\title{Unveiling Contrastive Learning's Capability of Neighborhood Aggregation for Collaborative Filtering}


\author{Yu Zhang}
\affiliation{%
  \institution{Anhui University}
  \city{Hefei}
  \country{China}
}
\email{zhangyu.ahu@gmail.com}

\author{Yiwen Zhang}
\authornote{Yiwen Zhang is the corresponding author.}
\affiliation{%
  \institution{Anhui University}
  \city{Hefei}
  \country{China}
}
\email{zhangyiwen@ahu.edu.cn}

\author{Yi Zhang}
\affiliation{%
  \institution{Anhui University}
  \city{Hefei}
  \country{China}
}
\email{zhangyi.ahu@gmail.com}

\author{Lei Sang}
\affiliation{%
  \institution{Anhui University}
  \city{Hefei}
  \country{China}
}
\email{sanglei@ahu.edu.cn}

\author{Yun Yang}
\affiliation{%
  \institution{Swinburne University of Technology}
  \city{Melbourne}
  \country{Australia}
}
\email{yyang@swin.edu.au}

\renewcommand{\shortauthors}{Yu Zhang, Yiwen Zhang, Yi Zhang, Lei Sang, \& Yun Yang}

\begin{abstract}
Personalized recommendation is widely used in the web applications, and graph contrastive learning (GCL) has gradually become a dominant approach in recommender systems, primarily due to its ability to extract self-supervised signals from raw interaction data, effectively alleviating the problem of data sparsity. 
A classic GCL-based method typically uses data augmentation during graph convolution to generates more contrastive views, and performs contrast on these new views to obtain rich self-supervised signals. 
Despite this paradigm is effective, the reasons behind the performance gains remain a mystery. 
In this paper, we first reveal via theoretical derivation that the gradient descent process of the CL objective is formally equivalent to graph convolution, which implies that CL objective inherently supports neighborhood aggregation on interaction graphs. 
We further substantiate this capability through experimental validation and identify common misconceptions in the selection of positive samples in previous methods, which limit the potential of CL objective. 
Based on this discovery, we propose the Light Contrastive Collaborative Filtering (LightCCF) method, which introduces a novel neighborhood aggregation objective to bring users closer to all interacted items while pushing them away from other positive pairs, thus achieving high-quality neighborhood aggregation with very low time complexity.  
On three highly sparse public datasets, the proposed method effectively aggregate neighborhood information while preventing graph over-smoothing, demonstrating significant improvements over existing GCL-based counterparts in both training efficiency and recommendation accuracy. 
Our implementations are publicly accessible\footnote{\url{https://github.com/ZzYUuuu/LightCCF}}\footnote{\url{https://github.com/BlueGhostYi/ID-GRec}}.

\end{abstract}




\maketitle

\section{Introduction} 
Recommender systems \cite{Wu_SurveyRecSys_TKED_2023,gao_gnn2-survey_RS_2023} have become an essential part of providing personalized services on video platforms, e-commerce sites, and various other web applications \cite{Wu_SurveyRecSys_TKED_2023}. 
To effectively capture user preferences, collaborative filtering (CF) \cite{Koren_CF_survey_2022} approach is widely employed due to its simplicity and effectiveness. 
The core principle of CF is to predict similar items that users might appreciate based on their existing interactions. 
Recently, with the rise of contrastive learning (CL) \cite{jing_CL-survey_2023,yu_ssl-survey_TKDE_2024}, there has been an increasing number of researches \cite{Assran_ssl-image_CVPR_2023,Chen_Human-Visual_CVPR_2023} that are attempting to incorporate this paradigm. The key advantage of CL can extract self-supervised signals from sparse raw interaction data, effectively addressing the issue of data sparsity \cite{xia_AutoCF_WWW_2023,xu_LightGCL_ICLR_2023}. 
Meanwhile, in CF, data sparsity has been a persistent challenge, leading to the emergence of numerous graph contrastive learning (GCL)-based methods. 
For instances, SGL \cite{wu_SGL_SIGIR_2021} and SimGCL \cite{yu_SimGCL_SIGIR_2022} typically generate contrastive views via data augmentation ($e.g.,$ dropout node or edge and perturb embedding), which are subsequently leveraged in CL to obtain self-supervised signals for addressing data sparsity and enhancing recommendation performance.

Despite CL having achieved excellent results in CF, the reasons behind its performance improvement remain unclear. Most existing CL-based methods \cite{he_CGCL_SIGIR_2023,Lin_NCL_WWWW_2022} focus on finding more and more effective contrastive views, aiming to capture valuable self-supervised signals from these new views. 
However, there has been little research exploring the core of contrastive learning—the CL objective, particularly InfoNCE loss \cite{chen_InfoNCE_ICML_2020}. 

In an effort to uncover the rationale behind CL’s efficacy, we first explore the gradient descent process of InfoNCE (as detailed in Section \ref{sec:theoretical}). 
Through a series of derivations, we find that InfoNCE optimizes representation uniformity during gradient descent. 
However, this does not directly explain how CL loss improves recommendation performance. 
To further explore the underlying mechanisms of InfoNCE loss, we train methods without data augmentation, using common user-user and item-item positive pairs \cite{yu_SimGCL_SIGIR_2022,wu_SGL_SIGIR_2021}. We compare the performance with and without a GCN encoder. As discussed in Section \ref{sec:eff_info_gcn}, the results show that both CL loss and the GCN encoder are effective, and combining them significantly improves model performance. This raises a key question: \textit{Is the improvement due to the CL objective itself or merely the addition of the GCN encoder?}
We argue that the limited gains observed without a GCN encoder may be due to incorrect positive sample selection—the users' positive sample should be the items they interacted with, not themselves. To test this, we replace the positive pairs (user-item pairs) and revisit the gradient descent process of the CL loss. We find that the CL objective inherently exhibits the neighborhood aggregation ability typical of graph convolutions \cite{Liu_GCN_Survey_JAS_2022, Sun_Neighborhhod_AAAI_2020}. Similarly, to further validate the neighborhood aggregation capability of CL loss, we adjust the positive pairs selection to user-item pairs and retrain the model. Results show that even with a base encoder, the CL loss effectively achieves neighborhood aggregation.

From above experimental results, we uncover new insights into the role of CL objective. While CL is known for capturing diverse self-supervised signals, our findings suggest that its true strength lies in its potent neighborhood aggregation capability. When users or items pairs are treated as contrastive views using a base encoder, the performance improvement is significantly less compared to when a GCN encoder is employed. This disparity arises because the GCN encoder enriches the model with substantial neighborhood information, allowing the CL loss to effectively aggregate neighborhood features. 
Subsequently, when we redefine positive pairs as user-item pairs and apply the same base encoder, the model's performance surpasses that of these previously proposed methods. This compelling result highlights InfoNCE's inherent capability for neighborhood aggregation, even in the absence of a GCN encoder. 

In this paper, we present a novel method, called \textbf{Light} \textbf{C}ontrastive \textbf{C}ollaborative \textbf{F}iltering (\textbf{LightCCF}). It is designed to effectively aggregate neighborhood information while mitigating the risk of over-smoothing in interaction graphs. Recognizing the limitations of InfoNCE loss in fully leveraging neighborhood aggregation, we introduce a neighborhood aggregation loss function. This function optimizes learning by bringing users closer to all their interacted items and distancing them from other positive pairs, thereby enhancing mutual information between users and their interacted items for more effective neighborhood aggregation. 

The primary contributions of this paper are as follows:

\begin{itemize}[left=0pt]
    \item We theoretically and experimentally explain why contrastive learning improves recommendation performance and reveal that the InfoNCE loss possesses neighborhood aggregation capability.

    \item We propose a simple yet effective contrastive collaborative filtering (LightCCF) method that efficiently aggregates neighborhood information while preventing graph over-smoothing. This method offers new research directions for CL recommendation.

    \item We conduct comprehensive experiments on three highly sparse datasets, demonstrating that the proposed method significantly outperforms GCL-based methods in terms of recommendation accuracy and training efficiency.
\end{itemize}

\section{Preliminaries}




\subsection{Problem Definition} 
In the recommender system scenario, there usually exists a set of users $\mathcal{U}$ and a set of items $\mathcal{I}$, where $|\mathcal{U}| = M$ and $|\mathcal{I}| = N $ are the number of users and items, respectively. 
Subsequently, a user-item interaction matrix $\mathbf{R} \in \mathbb{R}^{M \times N}$ is obtained from the interaction behavior of users and items. 
Our aim is to utilize known interactions to infer unknown ones, allowing for the provision of personalized recommendations for each user. 

\subsection{Graph and Graph Convolution} 
Recent researches \cite{wang_GraphCF_SIGIR_2020,wang_NGCF_SIGIR_2019} have proposed the concept of graph in collaborative filtering, and based on this, a definition of neighborhood aggregation that attempts to capture neighborhood information through graph convolution. 
Specifically, the recommendation task presents a user-item bipartite graph \cite{fan_GraphRec_WWW_2019} in terms of graph modeling: $\mathcal{G}=\langle\mathcal{V}=\{\mathcal{U},\mathcal{I}\}, \mathcal{E}\rangle$, where $\mathcal{V} = \{u_1,...,u_m,i_1,...,i_n \}$ denotes all user and item nodes, and $\mathcal{E} = \{u_1 \rightarrow i_1,..., u_m \rightarrow i_n\}$ denotes the interaction edges, and the adjacency matrix is 
$\mathbf{A} \in \mathbb{R}^{(M+N) \times (M+N)}$. 
Graph structure excels in aggregating neighborhood information, mainly due to its graph filter $\mathrm{H}(\mathbf{L})$ \cite{Isufi_graph-signal_TSP_2024}, which effectively reduces high-frequency components that may correspond to noise or abrupt changes, while preserving low-frequency components that capture important global structures, as follow: 
\begin{equation}
    \mathrm{H}(\mathbf{L}) = \sum^{K}_{k=0} \alpha_k \mathbf{L}^k,
\end{equation}
where $\mathbf{L} = \mathbf{D} - \mathbf{A}$ is the Laplace matrix, $\mathbf{D}$ is the degree matrix of the adjacency matrix $\mathbf{A}$, $\alpha_k$ is the weight of $\mathbf{L}^k$. 
LightGCN \cite{he_LightGCN_SIGIR_2020}, widely used in recommender systems, adopts this approach by utilizing the normalized Laplacian matrix $\mathbf{\tilde{A}} = \mathbf{D}^{-\frac{1}{2}} \mathbf{A} \mathbf{D}^{-\frac{1}{2}}$ as a graph filter to equate the graph convolution operation. 
\begin{equation}
    \mathrm{H}'(\mathbf{\tilde{A}}) = \mathrm{H}(\mathbf{\tilde{A}}) x,
\end{equation}
where $\mathrm{H}'$ is the graph convolution filter, $x$ is the graph signal. 
Meanwhile, for efficient computation, we give the corresponding matrix computation procedure at the same time. 
\begin{align}
    \label{eq:gcn}
    \mathbf{E}^{(l+1)} 
      = \alpha_l \mathbf{\tilde{A}} \mathbf{E}^{(l)}
      = \prod_{j=0}^{l} \alpha_j \mathbf{\tilde{A}}^j \mathbf{E}^{(0)},
\end{align}
where $l$ is the number of GCN layers, $\alpha$ is the weighting for the different layers, and $\mathbf{E}^{(0)}$ is the base encoder. 

\subsection{Contrastive Learning} 
\label{sec:CL}
In recent years, contrastive learning (CL) \cite{He_InfoNCE_CVPR_2020,chen_InfoNCE_ICML_2020} has been widely adopted due to its powerful ability to mitigate data sparsity. The key to CL is to select positive and negative samples. In natural language processing \cite{gao_simcse_arixv_22}, positive samples are typically generated from the same sentence information through different neural network, while negative samples are derived from different sentences information.

However, in recommender systems, positive samples can be obtained in two primary ways: one approach involves applying perturbations to the embedding of the same sample via data augmentation to generate positive pairs; the other approach directly considers user-item interactions as positive pairs. Inspired by advances in other domains such as computer vision (CV) \cite{Chen_Human-Visual_CVPR_2023} and natural language processing (NLP) \cite{gao_simcse_arixv_22}, existing methods predominantly adopt data augmentation techniques to acquire positive pairs, with InfoNCE serving as the primary contrastive loss function: 

\begin{equation}
\label{eq:InfoNCE}
    l_{\mathrm{InfoNCE}}(i,j) = -\mathrm{log} 
    \frac
    {\mathrm{exp}(\mathrm{sim}(\mathbf{e}_i,\mathbf{e}_j)/ \tau)}
    {\sum_{k \in \mathcal{K}} \mathrm{exp}(\mathrm{sim}(\mathbf{e}_i,\mathbf{e}_k) / \tau)},
\end{equation}
where \(i\) and \(j\) denote positive pairs ($i.e.,$ $\langle i,j \rangle$), and \(\mathcal{K}\) represents batch negative pairs, and $\tau$ is the temperature coefficient, and $\mathrm{sim}(\cdot)$ measures the embedding similarity. 
The objective of the InfoNCE loss is to draw positive pairs closer together while simultaneously distancing them from all negative pairs. In this way, it maximizes the mutual information between positive pairs, providing effective self-supervised signals to alleviate the challenge of data sparsity. 

\section{Investigation of Contrastive Learning in Recommendation}
In this section, we derive the form of gradient descent for CL objective and demonstrate InfoNCE's neighborhood aggregation capability from both theoretical and experimental perspectives.

\subsection{Exploring InfoNCE Gradient Descent}
\label{sec:theoretical}
Given the widespread application of the contrastive learning in recommender system \cite{wu_SGL_SIGIR_2021,yu_SimGCL_SIGIR_2022}, we focus on further exploring the core contrastive loss function widely adopted in these studies: InfoNCE loss \cite{He_InfoNCE_CVPR_2020}.
Specifically, we derive the gradients of the InfoNCE loss and introduce the gradient descent process as follows: 

\begin{align}
    \frac{\partial l_{\mathrm{InfoNCE}}(i,j)}{\partial \mathbf{e}_i} &= 
    -\frac{1}{\tau} \left(\mathbf{e}_j - \sum_{k \in \mathcal{K}} p_{ik} \mathbf{e}_k \right) , 
\end{align}
where $p_{ik}$ represents the softmax probability of the $k$ th sample:
\begin{equation}
    p_{ik} = \frac
    {\mathrm{exp}(\mathbf{e}_i \cdot \mathbf{e}_k / \tau)}
    {\sum_{k' \in \mathcal{K}} \mathrm{exp}(\mathbf{e}_i \cdot \mathbf{e}_k' / \tau) }.
\end{equation}
The above formulas represent the gradient of the InfoNCE loss, reflecting the contribution of the positive pair \( \langle\mathbf{e}_i, \mathbf{e}_j \rangle \) and the negative pairs \( \langle\mathbf{e}_i, \mathbf{e}_k\rangle \) $w.r.t.$ \( \mathbf{e}_i \). 
After calculating the relevant gradients, the embedding vectors are typically updated using gradient descent: 

\begin{align}
    \mathbf{e}_{i}^{(t+1)} 
    & = 
    \mathbf{e}_{i}^{(t)}  - \eta \frac{\partial l_{\mathrm{InfoNCE}}(i,j)}{\partial \mathbf{e}_i} 
    \nonumber \\
    & = 
    \mathbf{e}_{i}^{(t)} + \frac{\eta}{\tau} \left(\mathbf{e}_j^{(t)} - \sum_{k \in \mathcal{K}} p_{ik} \mathbf{e}_k^{(t)} \right),
\end{align}
where $\eta$ denotes the learning rate, $t$ denotes the $t$-th gradient descent. In the above, we derive the general form of InfoNCE gradient descent. In the recommendation scenario, we replace the generic form with user pairs ($e.g.,$ $\langle u',u'' \rangle$) as positive pairs:
\begin{align}
\label{eq:rec_info_gd}
    \mathbf{e}_{u'}^{(t+1)} 
    =&
    \mathbf{e}_{u'}^{(t)} + \frac{\eta}{\tau} \left(\mathbf{e}_{u''}^{(t)} - \sum_{k \in \mathcal{K}} p_{u'k} \mathbf{e}_k^{(t)} \right),
    \nonumber\\
    =&
    \mathbf{e}_{u'}^{(0)} + \sum_{i=0}^{t-1} \frac{\eta}{\tau} \left( \mathbf{e}_{u''}^{(i)} - \sum_{k \in \mathcal{K}} p_{u'k} \mathbf{e}_k^{(i)} \right),
\end{align}
where $\mathbf{e}_u'$ and $\mathbf{e}_u''$ are the embedding of user $u$ from different views, $\mathbf{e}_k$ are the negative pairs embedding, and $t \geq 1 $. 
Here, the term $\sum_{k \in \mathcal{K}} p_{u'k} e_k^{(t)}$ ensures that the distribution of the user set is well captured across the entire representation space \cite{wang_AU_ICML_2020}. 
After the training round, user \( u \) effectively uniformed the distribution of all negative pairs. 
The set of negative pairs is defined as $ U = \{ \{u_1, ..., u_{m-3}\},..., \{u_0, ..., u_{m-1}\} \} $, consisting of embeddings sampled from each batch. Ultimately, the uniform distribution of users is nearly achieved across all users ($U \approx \mathcal{U}$). 

\subsection{Effectiveness of InfoNCE and GCN Encoder}
\label{sec:eff_info_gcn}
Numerous methods leverage data augmentation to capture self-supervised signals and achieve a more uniform sample distribution \cite{yu_SimGCL_SIGIR_2022}. However, we aim to investigate the intrinsic properties of contrastive loss. To this end, we eschew commonly adopted data augmentation techniques and focus on analyzing contrastive loss in its original formulation. 
Specifically, we avoid generating contrastive views and GCN encoder for user \( u \) and instead directly compare identical user pairs \( \langle u, u \rangle \) and item pairs \( \langle i, i \rangle \), as follows: 
\begin{align}
    \mathcal{L}_{CL} 
    = -\mathrm{log} 
    \frac
    {\mathrm{exp}(\mathrm{sim}(\mathbf{e}_u,\mathbf{e}_u)/ \tau)}
    {\sum_{u' \in \mathcal{B}} \mathrm{exp}(\mathrm{sim}(\mathbf{e}_u,\mathbf{e}_u') / \tau)} 
    -\mathrm{log} 
    \frac
    {\mathrm{exp}(\mathrm{sim}(\mathbf{e}_i,\mathbf{e}_i)/ \tau)}
    {\sum_{i' \in \mathcal{B}} \mathrm{exp}(\mathrm{sim}(\mathbf{e}_i,\mathbf{e}_i') / \tau)},
\end{align}
where $\mathcal{B}$ represents the set of users and items in a batch, and the CL losses are based on Equation \ref{eq:InfoNCE}. 
We conduct the performance comparison experiments on two standard benchmark datasets: \textit{Douban-book} and \textit{Tmall} \cite{yu_SimGCL_SIGIR_2022,ren_DCCF_SIGIR_2023} (more experimental details in Section \ref{sec:experimental}). The corresponding experimental results are shown in Table \ref{tab:cl_ne} under \textbf{CL-SS (BE)}, where SS denotes positive pairs derived from `self-samples' (e.g., \(\langle u_0, u_0 \rangle\)), BE denotes `base encoder'. 

\begin{table}
    \centering
    \caption{Performance comparison of various CL-based methods and CL objective variants.}
    \label{tab:cl_ne}
    \begin{tabular}{p{1.7cm}|>{\centering\arraybackslash}p{1.25cm}>{\centering\arraybackslash}p{1.25cm}|>{\centering\arraybackslash}p{1.25cm}>{\centering\arraybackslash}p{1.25cm}}
    \specialrule{0.75pt}{0pt}{0pt}
    \multirow{2}{*}{Method} & \multicolumn{2}{c|}{Douban-book} & \multicolumn{2}{c}{Tmall}  \\
    \cline{2-5} 
    & Recall@20 
    & NDCG@20  
    & Recall@20 
    & NDCG@20 \\
    \hline
    \hline
    BPR-MF
    &0.1292 &0.1147	
    &0.0547	&0.0400		 \\
    LightGCN	
    &0.1504	&0.1404		
    &0.0711	&0.0530		 \\
    CL-SS (BE)
    &0.1478	&0.1433		
    &0.0565	&0.0412		 \\
    CL-SS (GCN)
    &0.1648	&0.1595
    &0.0746	&0.0554		 \\

    \specialrule{0.75pt}{0pt}{0pt}
    \end{tabular}
\end{table}

As can be observed, \textbf{CL-SS (BE)} consistently outperforms the baseline method BPR-MF \cite{koren_mf_article_2009} without leveraging the GCN encoder and data augmentations, relying solely on CL loss. This highlights the effectiveness of CL loss in boosting recommendation performance by refining the representation space and ensuring a more uniform sample distribution. 
However, compared to LightGCN \cite{he_LightGCN_SIGIR_2020}, its performance remains suboptimal. 
We conjecture that the base encoder lacks sufficient neighborhood information, limiting the contrastive loss from obtaining richer self-supervised signals, which impacts the overall recommendation performance. 
To validate our conjecture, we incorporate a GCN encoder into the \textbf{CL-SS (BE)} and present the experimental results in \textbf{CL-SS (GCN)}. 
The results clearly show a significant performance improvement after adding the GCN encoder. This finding suggests that, with the GCN encoder, CL can further enrich the self-supervised signals and explains the widespread use of GCN encoders in current CL-based methods ($i.e.,$ contain rich neighborhood information). 
We believe the key reason lies in the additional neighborhood information provided by the GCN encoder, which allows InfoNCE loss to better optimize the distribution of the representation space, leading to more uniform distribution. 
However, we are still uncertain whether this significant improvement is primarily driven by the GCN encoder, the InfoNCE loss, or the combined effect of both.

\subsection{Neighborhood Aggregation in InfoNCE} 
Based on the discussion of positive samples in Section \ref{sec:CL}, we observe that existing methods are influenced by prior work, often selecting positive pairs from the same sample while neglecting the intrinsic positive pairs ($i.e.,$ user-item pairs) from recommender systems. 
However, the selection of positive pairs is crucial for the effectiveness of the CL loss. While we show the InfoNCE's potential in Section \ref{sec:eff_info_gcn}, incorrect selection of positive pairs may lead to an over-reliance on neighborhood information from the GCN encoder, potentially resulting in the over-smoothing in graph \cite{wu_SCCF_KDD_2024}. 

Shifting our focus back to recommendation task, we recognize that users exhibit clear item preferences. Specifically, the items that a user interacts with can be seen as the user's neighbors. Consequently, positive sample for users should not be self-referential but should consist of the items with which they interact, forming user-item pairs. 
Formally, we provide the gradient descent formulation of $l_{\mathrm{InfoNCE}}(u,i)$ based on Equation \ref{eq:rec_info_gd}:
\begin{align}
\label{eq:ui_gd}
    \mathbf{e}_{u}^{(t+1)} 
    =&
    \mathbf{e}_{u}^{(t)} + \frac{\eta}{\tau} \left(\mathbf{e}_{i}^{(t)} - \sum_{k \in \mathcal{K}} p_{uk} \mathbf{e}_k^{(t)} \right),
\end{align}
where $\mathbf{e}_u$ and $\mathbf{e}_i$ are the embeddings of the user and item that form user-item positive pairs based on the users' interaction behavior. 
To our surprise, unlike previous gradient forms, \(\mathbf{e}_u^{(t+1)}\) optimizes the uniformity of the user representation distribution during each gradient descent step and directly aggregates neighborhood information from the first-order neighbors \(\mathbf{e}_i^{(t)}\). 
After several rounds of gradient descent, the user embeddings progressively aggregate neighborhood information through the InfoNCE loss, refining the representation space and enhancing uniformity. 
Based on this process, we demonstrate that InfoNCE has the ability to aggregate neighborhood information. Although this process differs from the aggregation strategy used in graph convolutions:

\begin{table}
    \centering
    \caption{Performance comparison of various CL-based methods and user-item pairs CL objectives.}
    \label{tab:cl-ui(BE)}
    \begin{tabular}{p{1.7cm}|>{\centering\arraybackslash}p{1.25cm}>{\centering\arraybackslash}p{1.25cm}|>{\centering\arraybackslash}p{1.25cm}>{\centering\arraybackslash}p{1.25cm}}
    \specialrule{0.75pt}{0pt}{0pt}
    \multirow{2}{*}{Method} & \multicolumn{2}{c|}{Douban-book} & \multicolumn{2}{c}{Tmall}  \\
    \cline{2-5} 
    & Recall@20 
    & NDCG@20  
    & Recall@20 
    & NDCG@20 \\
    \hline
    \hline
    LightGCN	
    &0.1504	&0.1404		
    &0.0711	&0.0530		 \\
    SGL
    &0.1633	&0.1585		
    &0.0738	&0.0556		 \\
    NCL
    &0.1647	&0.1539		
    &0.0750	&0.0553		 \\
    SCCF
    &0.1711	&0.1639		
    &0.0772	&\textbf{0.0580}		 \\
    CL-UI (BE)
    &\textbf{0.1746} &\textbf{0.1691}
    &\textbf{0.0758} &0.0568		 \\

    \specialrule{0.75pt}{0pt}{0pt}
    \end{tabular}
\end{table}

\begin{align}
    \mathbf{e}_{u}^{(k+1)} = \sum_{i \in \mathcal{N}_u} \frac{1}{\sqrt{|\mathcal{N}_u|}\sqrt{|\mathcal{N}_i|}}\mathbf{e}_{i}^{(k)},
\end{align}
where $\mathcal{N}_u$ and $\mathcal{N}_i$ denote the user and item first-order neighbors, respectively. Both approaches share the common goal of aggregating neighborhood information.

To further demonstrate InfoNCE's capability for neighborhood aggregation, we calculate the InfoNCE loss using user-item pairs, termed \textbf{CL-UI (BE)} in Table \ref{tab:cl-ui(BE)} (UI represents `user-item pairs'): 
\begin{equation}
    \mathcal{L}_{CL} = -\mathrm{log} 
    \frac
    {\mathrm{exp}(\mathrm{sim}(\mathbf{e}_u,\mathbf{e}_i)/ \tau)}
    {\sum_{i' \in \mathcal{B}} \mathrm{exp}(\mathrm{sim}(\mathbf{e}_u,\mathbf{e}_{i'}) / \tau)}.
\end{equation}
Note that we do not employ the GCN encoder; instead, we utilize the base encoder to more directly demonstrate the neighborhood aggregation capability of InfoNCE. As shown in the experimental results presented in Table 2, the performance exhibits a significant improvement compared to LightGCN, even surpassing the latest baselines \cite{wu_SGL_SIGIR_2021,Lin_NCL_WWWW_2022,wu_SCCF_KDD_2024} on certain datasets. These findings indicate that, despite the absence of the GCN encoder, InfoNCE is still able to effectively aggregate neighborhood information, thereby enhancing recommendation quality. This experimental outcome, along with the previous analysis, effectively addresses the questions raised in the preceding section and highlights the crucial role of InfoNCE in neighborhood aggregation and model performance improvement.

\section{Methodology}
The previous section demonstrated that the capability of InfoNCE for neighborhood aggregation. Building on this finding, our goal is to develop a light yet effective method that leverages CL loss for neighborhood aggregation in recommendation. This method offers several advantages: more interpretable, more effective at aggregating neighborhood information compared to GCN, and better at preventing the issue of over-smoothing, etc. 

\subsection{LightCCF}
In the base InfoNCE loss, although we identify appropriate positive sample, the sampling of negative samples remains a challenging issue. 
In recommender systems, most researchers use the InfoNCE loss as CL objective yet rarely modify the negative sampling process. Existing methods typically perform CL from multiple perspectives to obtain diverse self-supervised signals, but often overlook the contribution of negative pairs to the mutual information between positive pairs. 
To fully leverage the potential of InfoNCE, we present the \textbf{L}ight \textbf{C}ontrastive \textbf{C}ollaborative \textbf{F}iltering (\textbf{LightCCF}) method, which maximizes mutual information with interacted items by distancing the current user from other user-item positive pairs, thereby achieving more effective neighborhood aggregation. 

\begin{figure}[!t]
  \centering
  \includegraphics[width=0.85\linewidth]{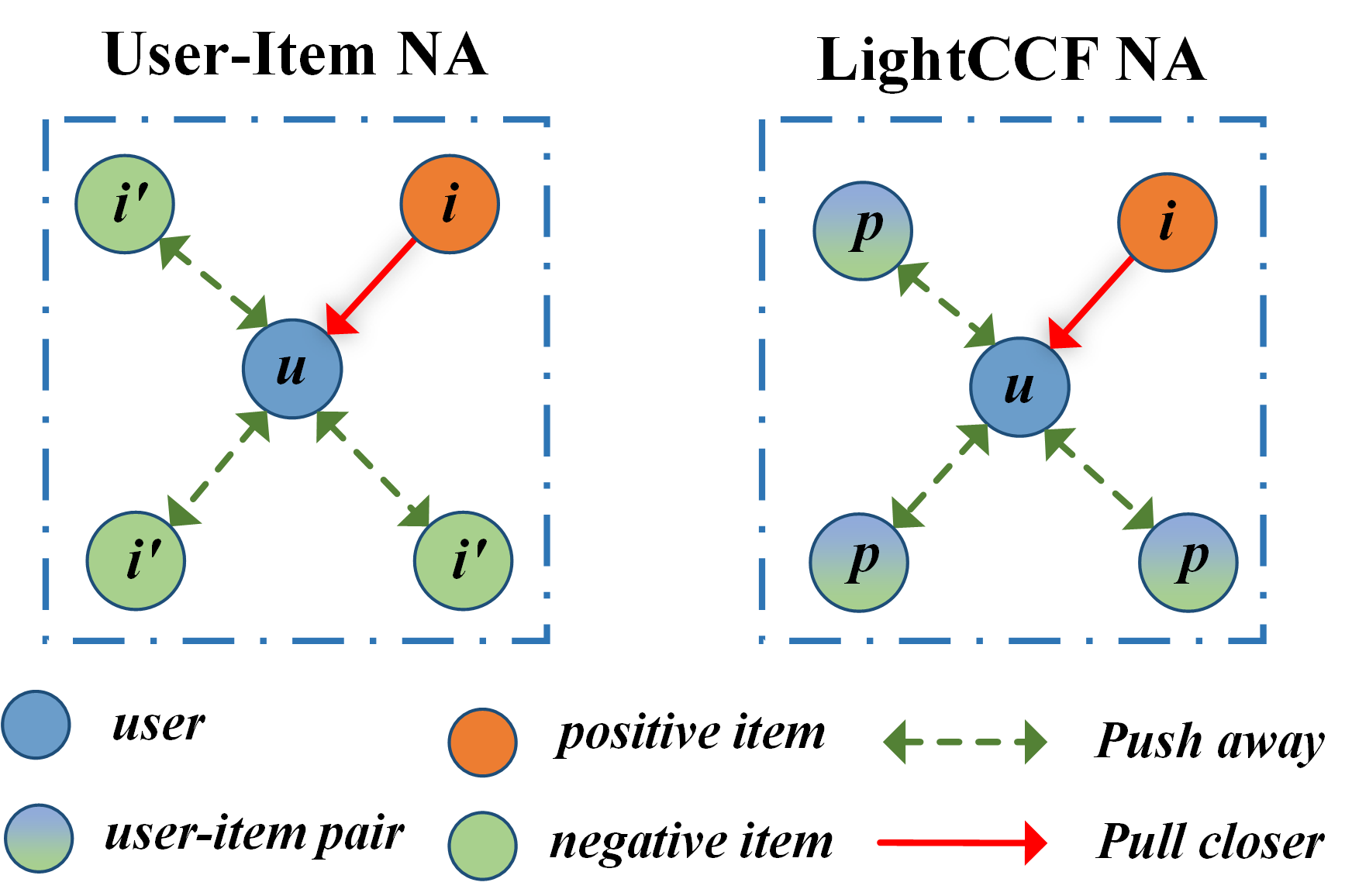 }
  \caption{Sampling strategies for neighborhood aggregation (NA) loss. Left: positive pairs are user-item pairs, and negative pairs are the item set. Right: positive pairs are user-item pairs, and negative pairs are other user-item pairs.}
  \label{fig:sampling}
\end{figure}

Formally, given a user $u$ and the interacted item set, we identify a set of positive pairs embedding for user $u$: 
$\mathcal{P}_{u} = \{ \mathbf{e}_u \rightarrow \mathbf{e}_{i \in \mathcal{N}_{(u)}} \}, $
where $\mathcal{N}_{(u)}$ is the user's first-order item neighbors. 
And the remaining positive pairs, excluding the current user, are considered as negative pairs: $\mathcal{P}/\{u\} = \{ \mathbf{e}_{u'} \rightarrow \mathbf{e}_{i \in \mathcal{N}_{(u')}} \}.$ We propose the following learning objective:

\begin{equation}
\label{eq:NA}
    \mathcal{L}_{NA} = \sum_{i \in \mathcal{N}_{(u)}} (
                       - \mathrm{sim}(\mathbf{e}_u,\mathbf{e}_i)/\tau 
                       + \mathrm{log}\sum_{p \in \mathcal{P}/\{u\}} \mathrm{exp}(\mathrm{sim}(\mathbf{e}_u,\mathbf{e}_{p})/\tau)),
\end{equation}
where $p$ is other positive pair. Specifically, to further exploit the neighborhood aggregation capability of the InfoNCE loss, LightCCF introduces a neighborhood aggregation (NA) loss function. This loss function consists of two main components. 
\textbf{First part:} $\mathrm{sim}(\mathbf{e}_u, \mathbf{e}_i)$, is used to aggregate information from the user’s neighborhood items; 
\textbf{Second part:} $\mathrm{sim}(\mathbf{e}_u, \mathbf{e}_p)$, is used to separate the user from all other user-item pairs. 
Through these two components, LightCCF more effectively achieves neighborhood aggregation, thereby improving recommendation quality. In addition, to clearly demonstrate the sampling strategy of LightCCF, we present the User-Item NA ($i.e.,$ \textbf{CL-UI}) method and the LightCCF NA method in Figure \ref{fig:sampling}. 

Based on Equation \ref{eq:InfoNCE}, we design an NA loss function to replace GCN in aggregating neighborhood information. For efficient training, we select a subset of other positive pairs as negative samples for each user-item pair to guide the neighborhood aggregation. Consequently, the NA loss in Equation \ref{eq:NA} can be restated as follows:
\begin{equation}
    \mathcal{L}_{NA} = -\mathrm{sim}(\mathbf{e}_u, \mathbf{e}_i)/\tau + \mathrm{log} \sum_{p \in \mathcal{B}_{p}} \mathrm{exp}(\mathrm{sim}(\mathbf{e}_u,\mathbf{e}_p)/\tau),
\end{equation}
where $\mathcal{B}_p$ is the set of positive pairs under batch condition.

\begin{table*}
    \centering
    \caption{Time complexity comparison in LightCCF with the current state-of-the-art methods. }
    \label{tab:time_complexity}
    \setlength{\extrarowheight}{2pt} 
    \begin{tabular}{c|c|c|c|c|c|c}
        \specialrule{0.75pt}{0pt}{0pt}
        Component & LightGCN & SGL & SimGCL & BIGCF  & RecDCL & LightCCF \\
        \hline
        \hline
        Adjacency Matrix & $\mathcal{O}(2|\mathcal{E}|)$ & $\mathcal{O}(2|\mathcal{E}| + 4\rho|E|)$ & $\mathcal{O}(2|\mathcal{E}|)$ & $\mathcal{O}(2|\mathcal{E}|)$ & $\mathcal{O}(2|\mathcal{E}|)$ & - \\
        GCN Process & $\mathcal{O}(2|\mathcal{E}|Ld)$ & $\mathcal{O}((2 + 4\rho)|\mathcal{E}|Ld)$ & $\mathcal{O}(6|\mathcal{E}|Ld)$ & $\mathcal{O}(2|E|Ld+|K||V|d)$ & $\mathcal{O}(2|\mathcal{E}|Ld)$ & - \\
        BPR Loss & $\mathcal{O}(2\mathcal{B}d)$ & $\mathcal{O}(2\mathcal{B}d)$ & $\mathcal{O}(2\mathcal{B}d)$ & $\mathcal{O}(2\mathcal{B}d)$ & - & $\mathcal{O}(2\mathcal{B}d)$ \\
        Additional Loss & - & $\mathcal{O}(2(\mathcal{B}d + \mathcal{B}^2d))$ & $\mathcal{O}(2(\mathcal{B}d + \mathcal{B}^2d))$ & $\mathcal{O}(5(\mathcal{B}d + \mathcal{B}^2d))$ & $\mathcal{O}(3\mathcal{B}d^2 + 3\mathcal{B}^2+2\mathcal{B}d)$   & $\mathcal{O}(\mathcal{B}d + 2\mathcal{B}^2d))$ \\
        \specialrule{0.75pt}{0pt}{0pt}
    \end{tabular}
\end{table*}

\subsection{Multi-Task Joint Training}

Although the NA loss can effectively integrate neighborhood information, the lack of clear optimization signals may lead to suboptimal performance in executing the primary CF task. Following previous works \cite{yu_SimGCL_SIGIR_2022,yang_VGCL_SIGIR_2023}, we introduce the widely-used Bayesian personalized ranking (BPR) \cite{rendle_bpr_2009} loss as follows:
\begin{equation}
    \mathcal{L}_{main} = -\mathrm{log}(\mathrm{sim}(\mathbf{e}_u,\mathbf{e}_i)-\mathrm{sim}(\mathbf{e}_u,\mathbf{e}_j)).
\end{equation}
where \( (u, i, j) \) represents a BPR triplet, with \( u \) as the user, \( i \) as the positive item, and \( j \) as the negative item. 

Through multi-task joint learning, we use the BPR loss as the primary loss to guide the optimization of NA loss, thereby further improving recommendation quality as follows:
\begin{equation}
    \mathcal{L} = \mathcal{L}_{main} + \alpha\mathcal{L}_{NA} + \beta\|\mathbf{E}^{(0)}\|^2_2.
\end{equation}
where \(\alpha\) is the neighborhood aggregation weight, and \(\beta\) is the  regularization weight, and $\mathbf{E}^{(0)}$ is the base encoder.

\subsection{Discussion}

\subsubsection{Necessity of Graph Convolution}

In this section, we discuss the necessity of GCN and whether it is required in LightCCF. Specifically, especially LightGCN \cite{he_LightGCN_SIGIR_2020}, is widely adopted due to its simple structure and efficient node information propagation. As a result, many CL-based methods calculate the CL loss using a GCN encoder. Contrary to the traditional view that CL alleviates data sparsity by capturing self-supervised signals, we believe that the significant improvement in recommendation performance is mainly due to InfoNCE's strong neighborhood aggregation capability. This does not undermine the importance of GCN in recommender systems, but rather highlights its overuse, as neighborhood information is inherently difficult to capture. 

By researching the experimental results of various GCL-based methods, we observe that as the number of GCN layers increases, the performance gains become limited. As Wu $et al.$ \cite{wu_SCCF_KDD_2024} pointed out, repeated message passing in GCN often leads to over-smoothing in the graph. This phenomenon does not entirely hinder performance, but it causes user and item embeddings to include redundant neighborhood information, ultimately limiting recommendation effectiveness. 
Thus, we aim to replicate the benefits of the GCN encoder by employing the CL objective. Our proposed LightCCF introduces the NA loss, which effectively substitutes the GCN encoder, facilitating neighborhood aggregation while preserving graph smoothness. We acknowledge that while GCN encoders may cause over-smoothing, their utility remaines, particularly in datasets with many false-positive samples. In such cases, the GCN encoder can capture higher-order neighborhood information through non-continuous aggregation, making it a valuable component of LightCCF.

\subsubsection{Relation with SCCF} 

Most classic GCL-based methods (e.g., SGL \cite{wu_SGL_SIGIR_2021}, SimGCL \cite{yu_SimGCL_SIGIR_2022}, and VGCL \cite{yang_VGCL_SIGIR_2023}) primarily use InfoNCE for contrastive learning (CL). Thus, our research focuses on exploring contrastive learning based on the InfoNCE loss. In contrast, SCCF \cite{wu_SCCF_KDD_2024} analyzes the SSM loss, alignment loss, and uniformity loss as CL objectives. However, in existing GCL-based research, the application of the SSM loss is relatively limited, while alignment and uniformity losses mainly come from graph representation learning \cite{wang_AU_ICML_2020}. Although some studies \cite{yu_SimGCL_SIGIR_2022} suggest that CL can optimize uniformity metrics, there is still limited understanding of the specific relationship between them, and few studies explicitly treat alignment and uniformity losses as part of CL. Therefore, the derivations presented in SCCF do not fully and directly reveal the underlying mechanisms of CL, particularly the practical capabilities of the InfoNCE loss in recommender systems. 

In contrast, we directly investigate the widely used InfoNCE loss and explore its neighborhood aggregation ability from both theoretical and experimental perspectives. Based on this, we propose LightCCF. By maximizing the mutual information between a user and all their positive items, and further optimizing the uniformity of other user-item negative pairs, we fully leverage the benefits of CL in neighborhood aggregation. As a result, LightCCF outperforms current state-of-the-art baselines in both performance and efficiency.

\subsection{Method Analysis}
\subsubsection{Matrix Form Analysis}
To enhance training efficiency for LightCCF, we transform the gradient descent process of the contrastive learning objective, InfoNCE (Equation \ref{eq:ui_gd}), into matrix form, as follow:
\begin{align}
\label{eq:matrix_gradient_descent}
    \mathbf{E}^{(t+1)} = \mathbf{E}^{(t)} + \frac{\eta}{\tau}( \mathbf{Y}^{(t)} - \mathbf{P}\mathbf{E}^{(t)}),
\end{align}
where $\mathbf{E}$ and $\mathbf{Y}$ represent the matrix of users and the corresponding positive items, respectively, $\mathbf{P}$ denotes the users' softmax probability matrix for the items. To further explore its potential capability, we present how the embedding matrix \(\mathbf{E}^{(0)}\) evolves into \(\mathbf{E}^{(t+1)}\) via iterative gradient descent:

\begin{align}
    \mathbf{E}^{(t+1)} = \mathbf{L}' \mathbf{E}^{(t)} = \prod_{j=0}^{t} \mathbf{L}'^{j} \mathbf{E}^{(0)},
\end{align}
where $\mathbf{L}'$ comes from the user's probability distribution over the corresponding items:
\begin{align}
    \mathbf{L}' = \mathbf{I} + \frac{\eta}{\tau}(\mathbf{Y}/\mathbf{E} - \mathbf{P}).
\end{align}
It is well-known that GCN encoders construct an adjacency matrix from the user-item interaction graph and perform neighborhood aggregation using the graph Laplacian matrix. Although the adjacency matrix does not explicitly appear in the gradient descent process of the InfoNCE loss, we observe that the probability matrix (\(\mathbf{P}\)) and positive item matrix (\(\mathbf{Y}\)) also contain neighborhood information. Therefore, despite the different methods of embedding generation and gradient updates, we find that the matrix form of the InfoNCE gradient descent is essentially equivalent to the neighborhood aggregation process in graph convolution (Equation \ref{eq:gcn}), $i.e.,$ \( \mathbf{E}^{(t+1)} \approx \mathbf{E}^{(l+1)} \). This finding effectively demonstrates the powerful neighborhood aggregation capability of InfoNCE, which further improves recommendation performance.

\subsubsection{Time Complexity Analysis}

To demonstrate the efficiency of LightCCF, we investigate the time complexity. 
Table \ref{tab:time_complexity} shows the time complexity details of the state-of-the-art methods. The time complexity of LightCCF is main composed of two parts. The first part comes from the BPR loss in the main CF task, with a time complexity of \( \mathcal{O}(2\mathcal{B}d) \), where \( \mathcal{B} \) represents the batch size and \( d \) denotes the embedding dimension. The second part stems from the NA loss, which incurs an additional \( \mathcal{O}(\mathcal{B}^2d) \) time complexity over InfoNCE due to the use of other positive pairs as negatives. Thus, the overall time complexity of LightCCF is \( \mathcal{O}(3\mathcal{B}d + 2\mathcal{B}^2d) \). 
For all methods that adopt GCN encoders, to enable efficient training, the construction of the adjacency matrix requires a time complexity of \(\mathcal{O}(2|\mathcal{E}|)\), where \(|\mathcal{E}|\) is the number of observed interactions. Performing one graph convolution incurs a time complexity of \(\mathcal{O}(2|\mathcal{E}| L d)\), where \(L\) and \(d\) denote the number of GCN layers and the embedding dimension, respectively. 
In addition, for SGL, \(\rho\) represents the retention probability, and for BIGCF, \( |\mathcal{K}| \) denotes the number of collective intent nodes. \( |\mathcal{V}| = \mathcal{M} + \mathcal{N} \) represents the total number of user and item nodes. 
We observe the following insights: 
\begin{itemize}[left=0pt]
    \item Using base encoder for neighborhood aggregation in LightCCF results in a time complexity slightly higher than that of the MF-BPR method, which only uses the BPR loss.
    \item LightCCF demonstrates a more significant time complexity advantage over most GCL-based methods. 
    \item LightCCF reveals that the CL loss has the capability for neighborhood aggregation and introduces the NA loss to achieve high-quality neighborhood information. This discovery allows us to perform neighborhood aggregation even without using a GCN encoder, thereby significantly reducing the time complexity.
\end{itemize}

\begin{table}
  \caption{Statistics of the datasets.}
  \label{tab:dataset_statistics}
  \begin{tabular}{c|c|c|c|c}
    \bottomrule
    Dataset & \#Users & \#Items & \#Interactions & Density\\
    \hline
    \hline
    Douban-book & 13.0k & 22.3k & 792.1k & 0.27\% \\  
    \hline
    Tmall & 47.9k & 41.4k & 2619.4k & 0.13\%\\
    \hline
    Amazon-book & 52.6k & 91.6k & 2984.1k & 0.06\%\\

    \toprule
  \end{tabular}
\end{table}

\begin{table*}
    \centering
    \caption{A comparison between the proposed LightCCF method and state-of-the-art baselines is provided. The best values are shown in bold, while the second-best values are underlined. R@ refers to Recall@, and N@ refers to NDCG@. 'Improv.\%' indicates the relative improvement over the top-performing baseline. An asterisk (*) indicates significant improvements with a t-test \( p \) < 0.05 compared to the second-best baseline.}
    \label{tab:main_experiment}
    \begin{tabular}{ll|p{0.8cm}p{0.8cm}p{0.8cm}p{0.8cm}|p{0.8cm}p{0.8cm}p{0.8cm}p{0.8cm}|p{0.8cm}p{0.8cm}p{0.8cm}p{0.8cm}}
    \specialrule{0.75pt}{0pt}{0pt}
    \multicolumn{2}{c|}{Method} & \multicolumn{4}{c|}{Douban-book} &
    \multicolumn{4}{c|}{Tmall} & \multicolumn{4}{c}{Amazon-book} \\
    \cline{0-13} 
     Name & From'Year & R@10 & N@10 & R@20 & N@20 & R@10 & N@10 & R@20 & N@20 & R@10 & N@10 & R@20 & N@20 \\
    \hline
    \hline
    BPR-MF \cite{rendle_bpr_2009} & UAI'09 
    & 0.0849 & 0.1079 & 0.1292 & 0.1147 
    & 0.0312 & 0.0287 & 0.0547 & 0.0400 
    & 0.0170 & 0.0182 & 0.0308 & 0.0239 
    \\
    Mult-VAE \cite{liang_VAE_WWW_2018} & WWW'18
    &0.1156 &0.1532	&0.1670 &0.1604
    &0.0467	&0.0423 &0.0740	&0.0552 
    &0.0224	&0.0239 &0.0407	&0.0315	\\
    NGCF \cite{wang_NGCF_SIGIR_2019} & SIGIR'19 
    &0.0899 &0.1137 &0.1376 &0.1215		
    &0.0374 &0.0351 &0.0629 &0.0465
    &0.0199 &0.0200 &0.0337 &0.0262\\
    LightGCN \cite{he_LightGCN_SIGIR_2020} & SIGIR'20
    &0.1025 &0.1359 &0.1504 &0.1404		
    &0.0435 &0.0406 &0.0711 &0.0530
    &0.0228 &0.0241 &0.0411 &0.0315\\
    SGL \cite{wu_SGL_SIGIR_2021} & SIGIR'21 
    &0.1153 &0.1558 &0.1633 &0.1585 	
    &0.0457 &0.0434 &0.0738 &0.0556
    &0.0263 &0.0281 &0.0478 &0.0379\\
    NCL \cite{Lin_NCL_WWWW_2022} & WWW'22
    &0.1121 &0.1485 &0.1647 &0.1539 	
    &0.0459 &0.0429 &0.0750 &0.0553
    &0.0266 &0.0284 &0.0481 &0.0373\\
    SimGCL \cite{yu_SimGCL_SIGIR_2022} & SIGIR'22
    &0.1230 &0.1637 &\underline{0.1772} &0.1583 
    &\underline{0.0559} &\underline{0.0536} &\underline{0.0884}	&\underline{0.0674}
    &\underline{0.0313} &\underline{0.0334} &0.0515	&0.0414\\
    DirectAU \cite{wang_DirectAU_KDD_2022} & KDD'22
    &0.1153 &0.1527 &0.1660 &0.1568	
    &0.0475 &0.0443 &0.0752 &0.0576
    &0.0296 &0.0297 &0.0506 &0.0401 \\
    CVGA \cite{zhang_CVGA_TOIS_2023}& TOIS'23
    &0.1229 &\underline{0.1670}	&0.1763 &\underline{0.1699}
    &0.0540 &0.0517	&0.0854 &0.0648 
    &0.0290 &0.0302	&0.0492	&0.0379		\\
    LightGCL \cite{xu_LightGCL_ICLR_2023} & ICLR'23
    &0.1069 &0.1393 &0.1570 &0.1455 
    &0.0531 &0.0508 &0.0833 &0.0637
    &0.0303 &0.0318 &0.0506 &0.0397\\
    DiffRec \cite{wang_DiffRec_SIGIR_2023} & SIGIR'23
    &0.1102 &0.1545	&0.1619 &0.1661
    &0.0485 &0.0473	&0.0792 &0.0612
    &0.0310 &0.0333	&0.0514&\underline{0.0418}		\\
    CGCL \cite{he_CGCL_SIGIR_2023} & SIGIR'23
    &0.1216 &0.1622 &0.1741 &0.1667 
    &0.0542 &0.0510 &0.0880 &0.0655
    &0.0274 &0.0284 &0.0483 &0.0380\\
    VGCL \cite{yang_VGCL_SIGIR_2023} & SIGIR'23
    &0.1203 &0.1655 &0.1733 &0.1689
    &0.0557 &0.0533 &0.0880 &0.0670	
    &0.0312 &0.0332 &0.0515 &0.0410\\
    RecDCL \cite{Dan_RecDCL_WWW_2024} & WWW'24
    &0.1151 &0.1452 &0.1664 &0.1526 
    &0.0527 &0.0492 &0.0853 &0.0632	
    &0.0311 &0.0318 &\underline{0.0525} &0.0407\\
    BIGCF \cite{zhang_BIGCF_SIGIR_2024} & SIGIR'24
    &0.1199 &0.1642 &0.1741 &0.1682 
    &0.0547 &0.0524 &0.0876 &0.0664	
    &0.0294 &0.0320 &0.0500 &0.0398\\
    SCCF \cite{wu_SCCF_KDD_2024} & KDD'24
    &\underline{0.1252} &0.1611 &0.1711 &0.1639 
    &0.0478 &0.0453 &0.0772 &0.0580
    &0.0287 &0.0294 &0.0491 &0.0399\\
    \hline
    \hline
    LightCCF &  ---
    &\textbf{0.1330}* &\textbf{0.1848}* &\textbf{0.1877}* &\textbf{0.1857}*
    &\textbf{0.0599}* &\textbf{0.0574}* &\textbf{0.0949}* &\textbf{0.0724}*
    &\textbf{0.0347}* &\textbf{0.0375}* &\textbf{0.0577}* &\textbf{0.0462}* \\
    Improv.\% &  ---
    &6.23\% &10.66\% &5.93\% &9.30\%
    &7.16\% &7.09\% &7.35\% &7.42\%
    &10.86\% &12.28\% &9.90\% &10.53\% \\
    \specialrule{0.75pt}{0pt}{0pt}
    \end{tabular}
\end{table*}


\section{Experiments}
In this section, we conduct experimental comparisons with state-of-the-art recommendation methods on three real-world datasets to demonstrate the effectiveness of LightCCF.

\subsection{Experimental Settings}
\label{sec:experimental}

\subsubsection{Datasets.}
We select three widely used public benchmark datasets for our experiments: \textbf{Douban-book} \cite{yu_SimGCL_SIGIR_2022} contains book preference data from Douban, \textbf{Tmall} \cite{ren_DCCF_SIGIR_2023} includes online purchase records from the Tmall website, and \textbf{Amazon-book} \cite{yu_SUAU_ESWA_2024} consists of book purchase records from Amazon. The specific details of the datasets are provided in Table \ref{tab:dataset_statistics}. 

\subsubsection{Evaluation Indicators.} For all datasets, we follow previous works \cite{he_LightGCN_SIGIR_2020,zhang_CVGA_TOIS_2023} and randomly split each user's interactions into training and test sets with the 80\% and 20\% ratio, respectively. Additionally, to evaluate the Top-K recommendation performance of the LightCCF model compared to other models, we use the widely adopted metrics Recall@K and NDCG@K (K=10, 20).

\subsubsection{Baselines.} 
We compare LightCCF with several methods, which fall into four groups: 
\textbf{1) BPR- and AU-based methods}, which include randomly sampled negative instances and optimize the model with BPR loss ($e.g.,$ BPR-MF \cite{koren_mf_article_2009}), as well as those that optimize the model with alignment and uniformity losses ($e.g.,$ DirectAU \cite{wang_DirectAU_KDD_2022}); and 
\textbf{2) GCN-based methods}, which replace base encoder with a GCN encoder ($e.g.,$ NGCF \cite{wang_NGCF_SIGIR_2019} and LightGCN \cite{he_LightGCN_SIGIR_2020}); and 
\textbf{3) Generative-based methods}, which leverage generative models to learn the underlying distribution of user-item interactions ($e.g.,$ Mult-VAE \cite{liang_VAE_WWW_2018}, CVGA \cite{zhang_CVGA_TOIS_2023} and DiffRec \cite{wang_DiffRec_SIGIR_2023}); and 
\textbf{4) CL-based methods}, which incorporate CL to assist the recommendation task ($e.g.,$ SGL \cite{wu_SGL_SIGIR_2021}, NCL \cite{Lin_NCL_WWWW_2022}, SimGCL \cite{yu_SimGCL_SIGIR_2022}, LightGCL \cite{xu_LightGCL_ICLR_2023}, CGCL \cite{he_CGCL_SIGIR_2023},  VGCL \cite{yang_VGCL_SIGIR_2023}, RecDCL \cite{Dan_RecDCL_WWW_2024}, BIGCF \cite{zhang_BIGCF_SIGIR_2024} and SCCF \cite{wu_SCCF_KDD_2024}).

\subsubsection{Hyperparameters Detail.} To ensure a fair comparison, we identify the optimal parameters for each method based on the original papers and perform all experiments under the same conditions. We repeat each experiment five times and use the average results. For all methods, embeddings are randomly initialized using the Xavier \cite{glorot_Xavier_AI_2010} approach , with an embedding size of 64 (2048 for RecDCL), as RecDCL requires a larger embedding size due to its method design. The GCN encoder is configured with 1, 2, and 3 layers. 
The temperature coefficient \(\tau\) is set within \{0.20, 0.22, 0.24, 0.26, 0.28, 0.30\}, the NA loss weight $\alpha$ within \{0.1, 0.5, 1, 2.5, 5, 10\}, the regularization weight $\beta$ and learning rate are fixed at 0.0001 and 0.001, respectively. 

\subsection{Performance Comparison with Baselines}
To validate the effectiveness of LightCCF, we compare it against all baselines presented in Table \ref{tab:main_experiment}. Across all datasets, LightCCF consistently outperforms the baselines, achieving the best recommendation performance. Specifically, compared to the second-best baseline, LightCCF improves NDCG@10 by 11.66\%, 7.09\%, and 12.28\% on Douban-book, Tmall, and Amazon-book, respectively. Notably, while SimGCL and VGCL enhance embedding information through data augmentation and achieve strong results, our method without using such techniques yet still delivers superior recommendation quality. These improvements demonstrate that the NA loss in LightCCF effectively aggregates neighborhood information, leading to better recommendation performance.

The table shows that many existing baselines are CL-based methods, with most being GCL-based. This is mainly because the GCN encoder captures rich neighborhood information, allowing the CL loss to perform neighborhood aggregation and improve recommendation accuracy. However, since most existing methods treat users/items themselves as positive samples in their sampling strategies, CL loss can only aggregate neighborhood information from its own GCN encoder. Although these methods attempt to capture subtle differences in embeddings through data augmentation, the improvement in aggregation is limited. To address this challenge, LightCCF introduces a more adaptive NA loss that achieves comprehensive neighborhood aggregation and prevents graph over-smoothing even without graph convolution.

In addition, we also observe that AU-based methods demonstrate notable performance. These methods optimize the alignment and uniformity of representations through alignment and uniformity losses. However, since these methods only optimize alignment within positive pairs and uniformity within user and item representations, they overlook the connections between users and items. This results in a general lack of neighborhood information in such methods, making their recommendation performance inferior to LightCCF and even to some of the latest GCL-based approaches. 

\begin{table}
    \centering
    \caption{Performance comparison of LightCCF with and without GCN across different layers. Bold indicates optimal performance, and underline indicates sub-optimal performance. }
    \label{tab:encoder}
    \begin{tabular}{l|p{0.75cm}p{0.75cm}|p{0.75cm}p{0.75cm}|p{0.75cm}p{0.75cm}}
    \specialrule{0.75pt}{0pt}{0pt}
    \multirow{2}{*}{Method} & \multicolumn{2}{c|}{Douban-book} & \multicolumn{2}{c|}{Tmall} & \multicolumn{2}{c}{Amazon-book} \\
    \cline{2-7} 
    & R@20 
    & N@20 
    & R@20 
    & N@20  
    & R@20 
    & N@20 \\
    \hline
    \hline
    LightCCF-0
    &\underline{0.1866}	&\textbf{0.1857}		
    &\textbf{0.0949}	&\textbf{0.0724}
    &0.0542	&0.0422  \\
    \specialrule{0.25pt}{1pt}{1pt}
    LightCCF-1
    &\textbf{0.1878} &\underline{0.1854}
    &\underline{0.0917}	&\underline{0.0697}	
    &0.0562	&0.0437 \\
    LightCCF-2
    &0.1768	&0.1713	
    &0.0898	&0.0683
    &\underline{0.0565}	&\underline{0.0447} \\
    LightCCF-3
    &0.1780 &0.1724		
    &0.0902	&0.0686
    &\textbf{0.0577}	&\textbf{0.0462} \\
    \specialrule{0.75pt}{0pt}{0pt}
    \end{tabular}
\end{table}

\subsection{LightCCF based on GCN Encoder}
Although LightCCF achieves effective neighborhood aggregation with a base encoder, the rich neighborhood information in the GCN encoder is equally appealing to LightCCF. An important issue that emerges: \textit{Does incorporating the GCN encoder into LightCCF further enhance recommendation performance?} To explore this, we conduct experiments using GCN encoders with varying numbers of layers, with the results presented in Table \ref{tab:encoder}.

As shown in the table, most datasets do not see an improvement in performance after adding the GCN encoder; in fact, performance declines to some extent. We believe that this is due to the repeated neighborhood aggregation within the GCN encoder. Although the GCN encoder incorporates neighborhood information, the excess of this information can lead to over-smoothing in the graph, ultimately hindering recommendation performance. 
However, the results on the Amazon-book dataset are entirely different. As the number of GCN encoder layers increases, LightCCF's performance steadily improves. To explain this, we conduct a sparsity experiment in Section \ref{sec:sparsity}. Specifically, we find that for all baselines on Amazon-book, the performance of the sparse group is generally higher than that of the normal group, indicating that the normal group contains more false positive samples, making it harder for the model to capture user preferences. However, since graph convolution propagates information in a non-contiguous manner, the GCN encoder can effectively capture high-order neighborhood information, helping LightCCF mitigate the noise from false positive samples and achieve effective neighborhood aggregation.

\subsection{Data Sparsity Analysis}
\label{sec:sparsity}

\begin{figure}[!t]
  \centering
  \raisebox{2.25cm}{\rotatebox[origin=c]{90}{\small \textbf{NDCG@20(\%)}}}
  \hfill
  \begin{subfigure}[b]{0.31\linewidth}
    \includegraphics[width=\linewidth]{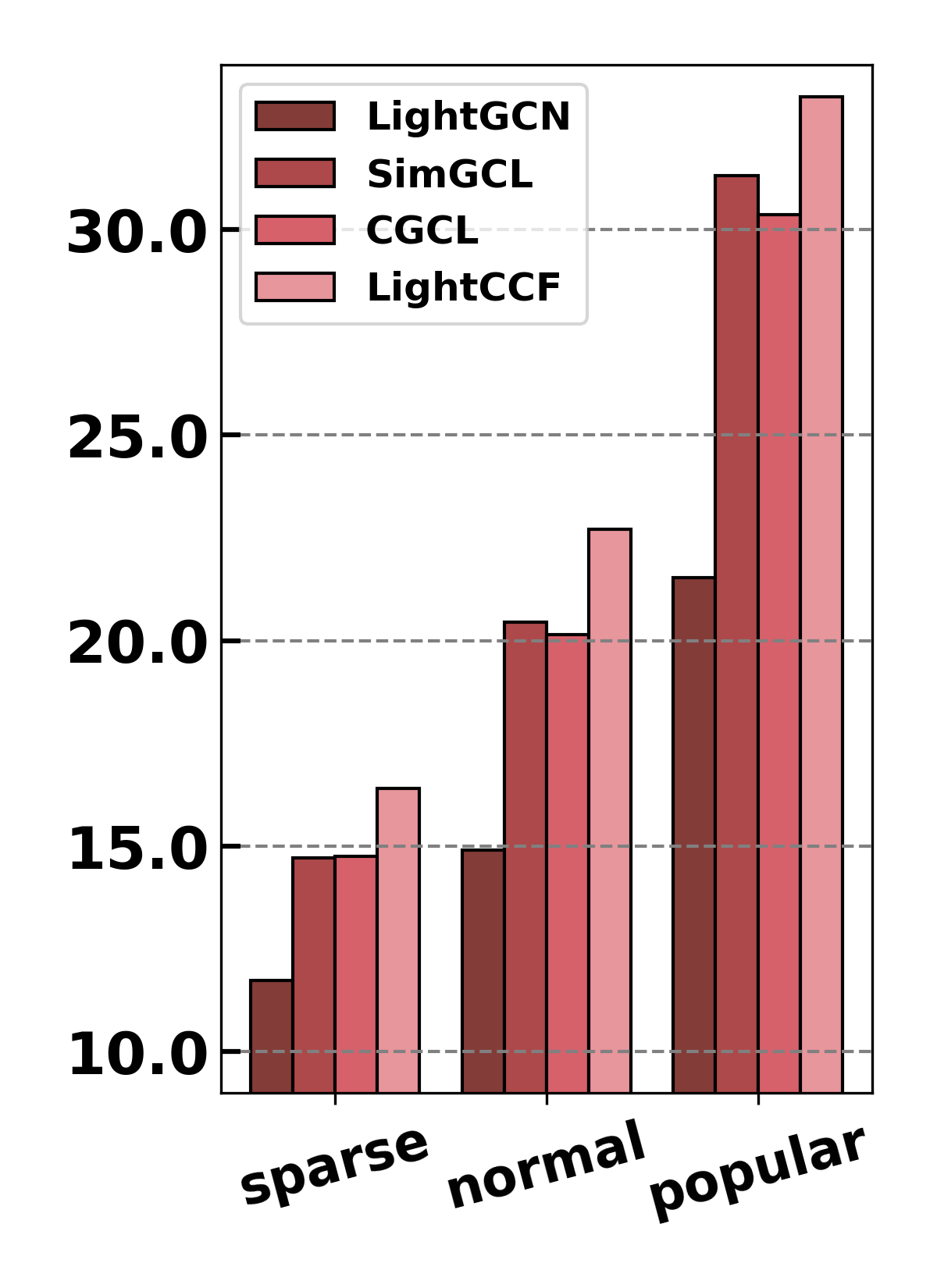}
    \caption{Douban-book}
    \label{fig:amazon-book_sparsity_ndcg@20}
  \end{subfigure}
  \hfill 
  \begin{subfigure}[b]{0.31\linewidth}
    \includegraphics[width=\linewidth]{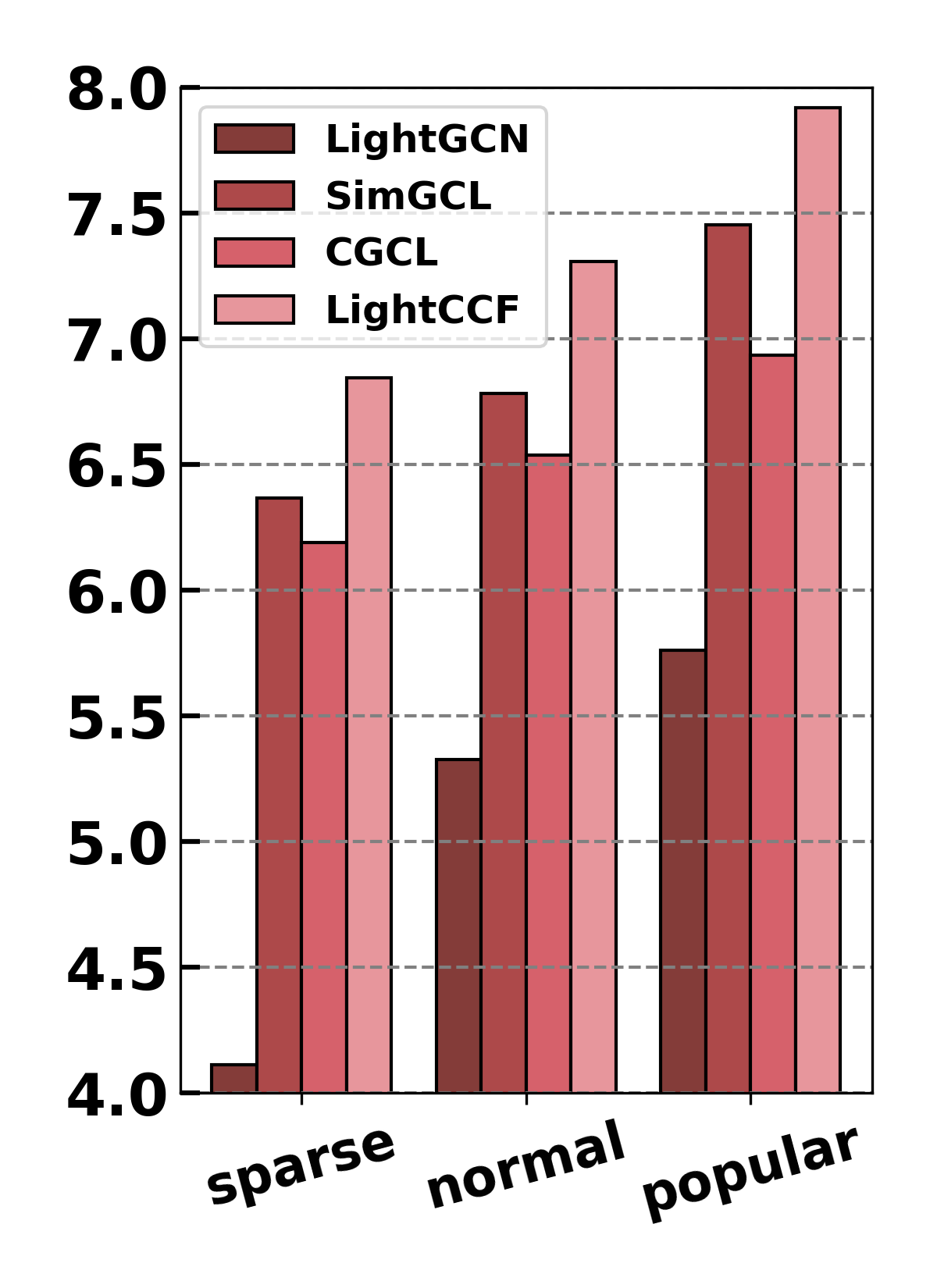}
    \caption{Tmall}
    \label{fig:yelp2018_sparsity_ndcg@20}
  \end{subfigure}
  \hfill 
  \begin{subfigure}[b]{0.31\linewidth}
    \includegraphics[width=\linewidth]{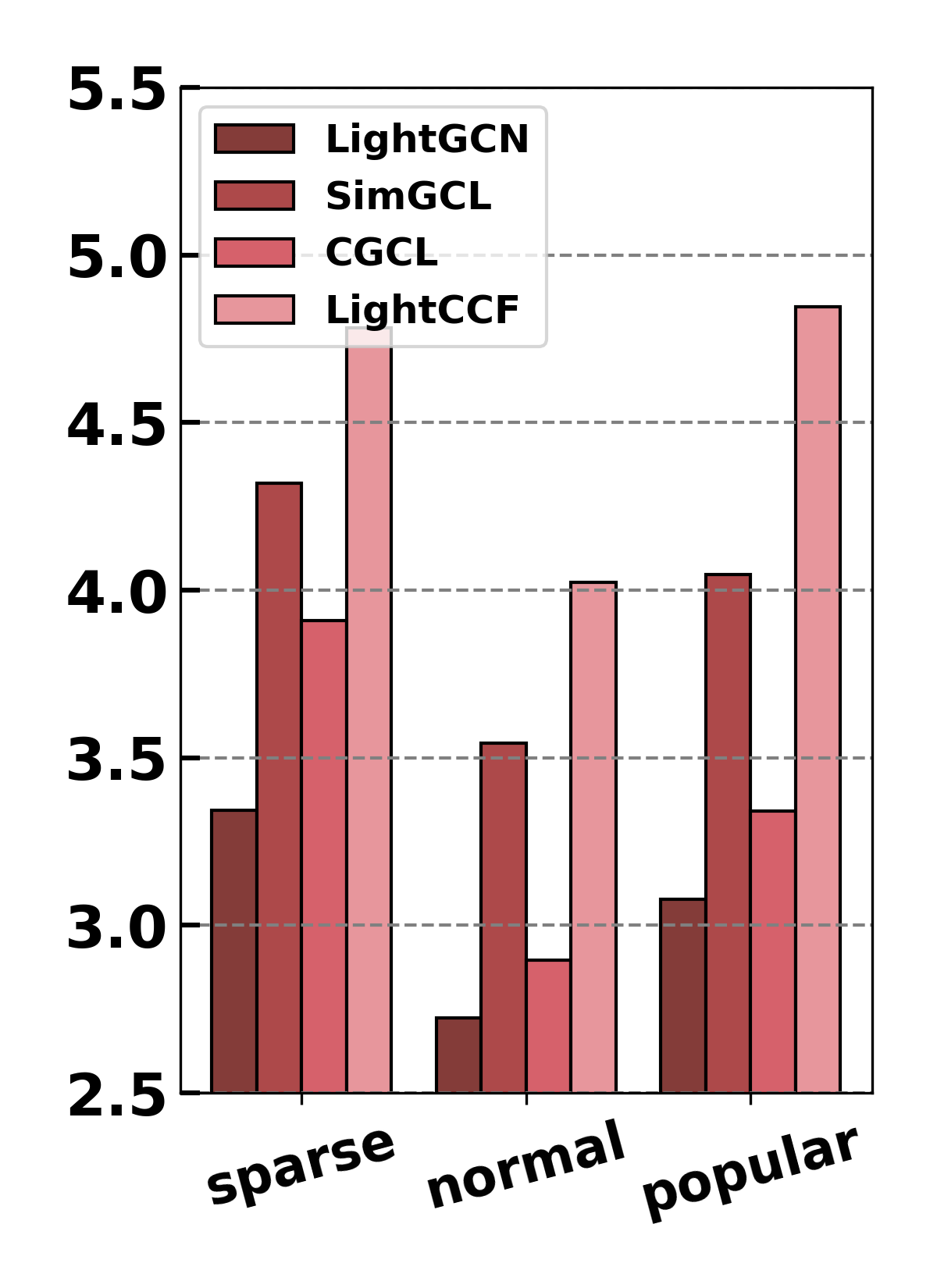}
    \caption{Amazon-book}
    \label{fig:tmall_sparsity_ndcg@20}
  \end{subfigure}
  \caption{Performance comparison \textit{w.r.t.} NDCG@20 of LightCCF and other models for different user groups sparsity levels on Douban-book, Tmall, and Amazon-book Datasets.}
  \label{fig:sparsity}
\end{figure}

\begin{table}
    \centering
    \caption{Efficiency comparison across two datasets, including per epoch time, number of epochs, and total runtime (measured in seconds (s), minutes (m), hours (h)). The best times are marked in bold for both per epoch and total runtime.}
    \label{tab:runtime}
    \begin{tabular}{l|ccc|ccc}
    \specialrule{0.75pt}{0pt}{0pt}
    \multirow{2}{*}{Method} & \multicolumn{3}{c|}{Douban-book} & \multicolumn{3}{c}{Tmall}  \\
    \cline{2-7} 
    & per 
    & epoch   
    & total
    & per   
    & epoch
    & total \\
    \hline
    \hline
    LightGCN
    &13.5s	&246 & 55m	
    &55.8s	&286 & 4h25m  
    \\
    SimGCL
    & 30.2s & 45 & 20m	
    & 142.4s & 24 & 56m	  
    \\
    LightGCL
    & 17.2s & 64 & 18m	
    & 65.1s & 52 & 56m	  
    \\
    CGCL
    & 22.6s & 76 & 28m 	
    & 134.1s & 76 & 2h49m
    \\
    VGCL
    & 20.1s & 47 & 16m 	
    & 74.5s & 52 & 1h5m
    \\
    SCCF
    & 10.2s & 71 & 12m	
    & 25.6s & 81 & 35m
    \\
    \hline
    LightCCF
    & \textbf{7.4s} & 41 & \textbf{5m} 	
    & \textbf{18.3s} & 80 & \textbf{18m}
    \\
    
    \specialrule{0.75pt}{0pt}{0pt}
    \end{tabular}
\end{table}

To evaluate whether LightCCF effectively alleviates the data sparsity issue, we compare it with the comparatively superior methods from Table \ref{tab:main_experiment}, as shown in Figure \ref{fig:sparsity}. Specifically, we divide the dataset into three groups—sparse, normal, and popular—based on the number of user interactions, and evaluate their performance $w.r.t.$ NDCG@20. 
The results indicate that CL-based methods outperform LightGCN, highlighting the effectiveness of contrastive learning in addressing data sparsity. LightCCF consistently outperforms the baselines due to its ability to leverage CL loss for neighborhood aggregation. 
Notably, on Amazon-book dataset, all methods perform worse in the normal group compared to other datasets, primarily due to the presence of more false positive samples. 
However, LightCCF still achieves the best performance, with its performance in the sparse group significantly surpassing the normal group, even matching the popular group. 

\subsection{Efficiency Analysis}


To test the training efficiency of LightCCF, we train all methods on a GeForce RTX 2080Ti GPU and list the actual runtimes in Table \ref{tab:runtime}. Most existing methods use a GCN encoder, which significantly increases the training time per epoch. However, the rich neighborhood information provided by GCN helps GCL-based methods converge faster, thereby reducing the overall runtime. LightCCF replaces GCN with the NA loss for neighborhood aggregation, resulting in higher training efficiency compared to the baselines. Notably, we exclude RecDCL’s results due to its large embedding size of 2048, which made it incompatible with the other methods on the same device and the most time-consuming method. 

\subsection{Robustness Analysis}



\begin{figure}[!t]
  \centering
  \hfill
  \begin{subfigure}[b]{0.49\linewidth}
    \includegraphics[width=\linewidth]{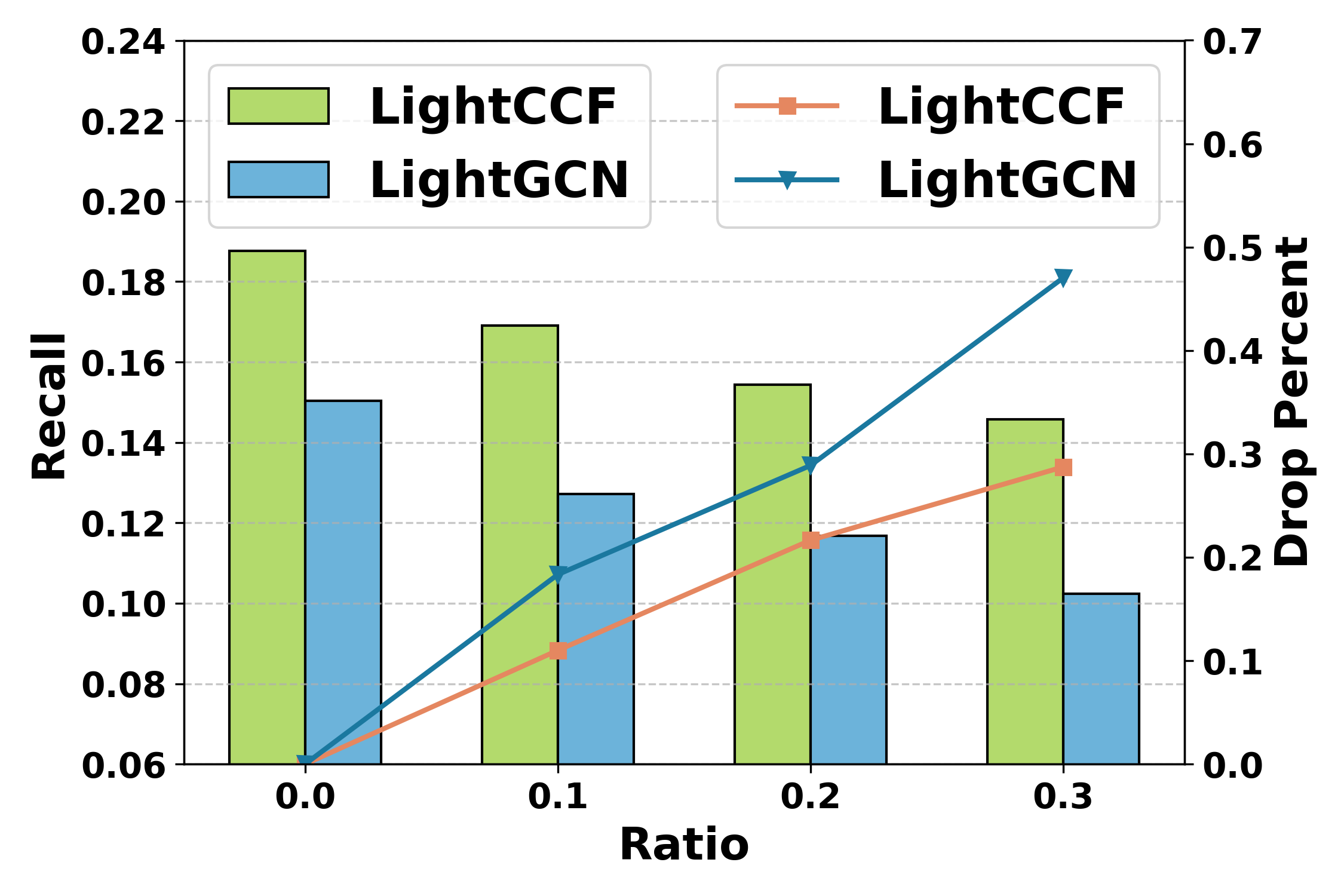}
    \caption{Douban-book}
  \end{subfigure}
  \hfill 
  \begin{subfigure}[b]{0.49\linewidth}
    \includegraphics[width=\linewidth]{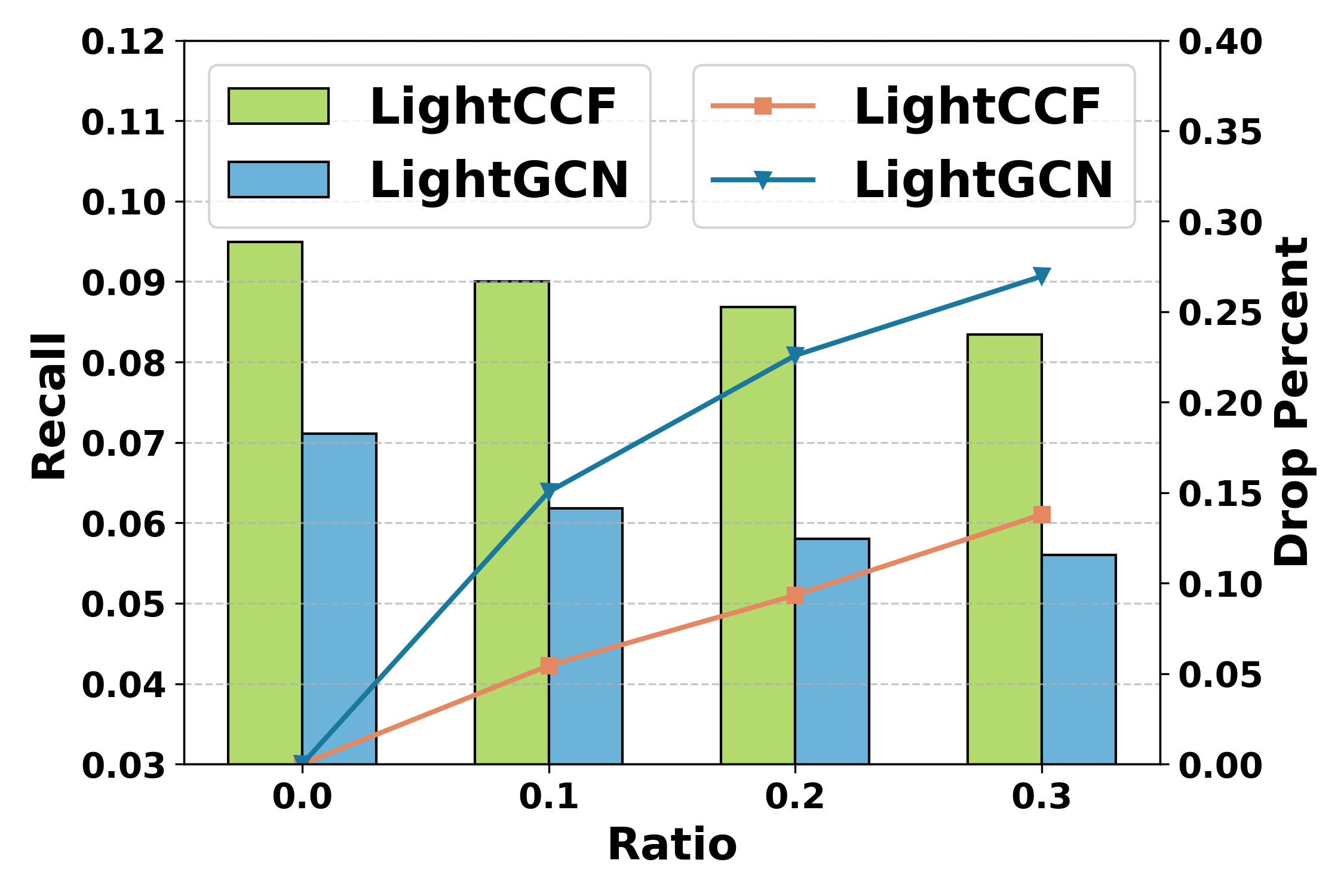}
    \caption{Tmall}
  \end{subfigure}
  \caption{Performance comparison \textit{w.r.t.} noise ratio. The bar represents Recall, while the line represents the percentage of performance degradation across two Datasets.}
\label{fig:rb}
\end{figure}

To evaluate the robustness of LightCCF under varying noise levels, we inject 10\%, 20\%, and 30\% adversarial user-item interactions into the training set while keeping the test set unchanged. As shown in Figure~\ref{fig:rb}, LightCCF maintains stable performance despite increasing noise, with only a slight decline in recall. Compared to LightGCN, it exhibits significantly lower performance degradation, demonstrating superior robustness. This strength stems from its innovative integration of user-item relationships and effective use of neighborhood information, which collectively reduce the impact of noise without compromising recommendation quality.

\subsection{Hyperparameter Analysis}

\begin{figure}[!t]
  \centering
  \hfill
  \begin{subfigure}[b]{0.49\linewidth}
    \includegraphics[width=\linewidth]{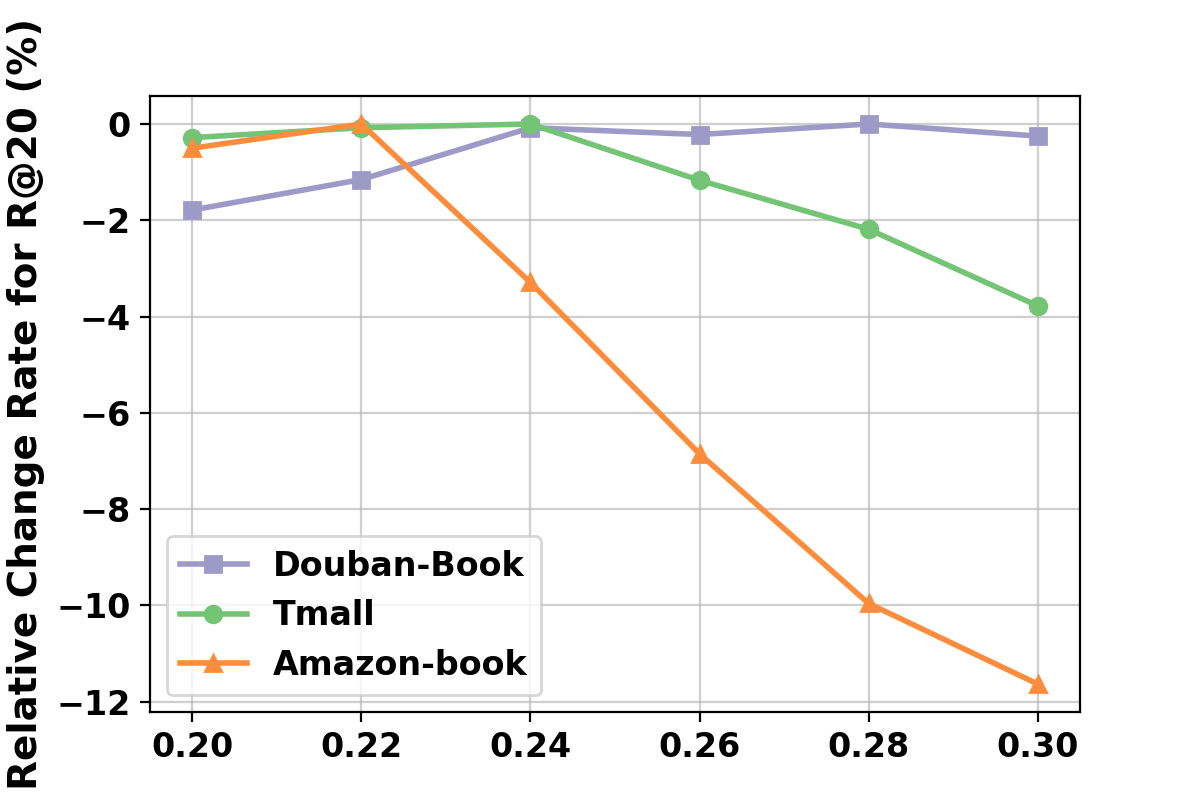}
    \caption{$\tau$}
  \end{subfigure}
  \hfill 
  \begin{subfigure}[b]{0.49\linewidth}
    \includegraphics[width=\linewidth]{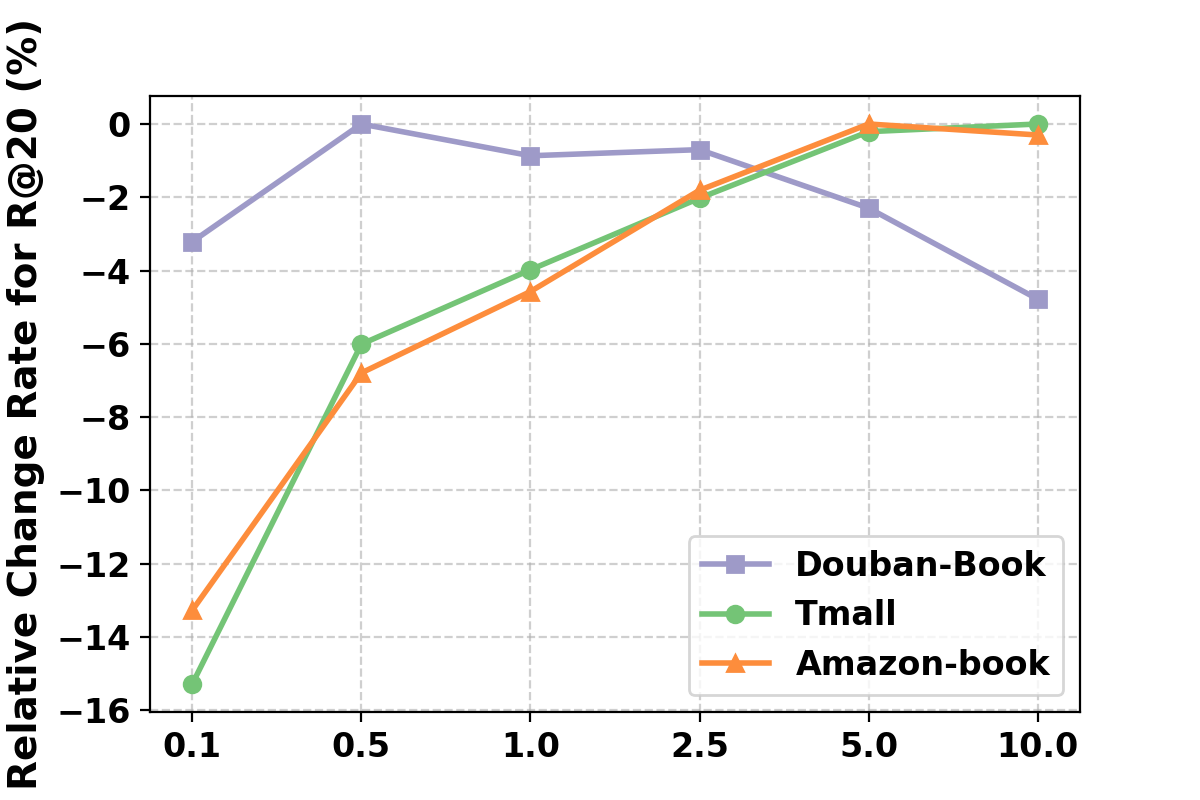}
    \caption{$\alpha$}
  \end{subfigure}
  \caption{Hyperparameter sensitivities to the temperature coefficient $\tau$ and the graph contrastive regularization weight $\alpha$ \textit{w.r.t.} Recall@20 across three datasets.}
  \label{fig:hyper}
\end{figure}

In this section, we explore the sensitivity of two key hyperparameters, as illustrated in Figure \ref{fig:hyper}. First, for the temperature coefficient \(\tau\). A smaller \(\tau\) heightens the influence of negative samples, particularly hard negatives, which contribute significantly to balancing mutual information between positive pairs.
As \(\tau\) increases, the contribution of individual samples is gradually replaced. 
Figure \ref{fig:hyper}(a) shows that because LightCCF treats other positive pairs as negative samples, the higher quality of these negatives leads to lower sensitivity to \(\tau\). 
Interestingly, on the Amazon-book dataset, \(\tau\) is notably lower, indicating that positive pairs in this dataset heavily rely on hard negatives, possibly due to a higher occurrence of false positives. 
Second, for the NA loss weight \(\alpha\), Figure \ref{fig:hyper}(b) depicts how varying weights affect performance. As \(\alpha\) increases, the performance generally improves across most datasets, allowing LightCCF to accurately capture neighborhood information. 
However, while higher \(\alpha\) boosts performance, it can also slow convergence. Therefore, an optimal \(\alpha\) is crucial for efficiently capturing user preferences, and delivering personalized recommendations.

\section{Related Work}

\medskip
\noindent\textbf{Graph Neural Networks in CF}. 
Collaborative filtering (CF) \cite{Koren_CF_survey_2022,su_CF_survey_2009} has become a cornerstone in recommender systems due to its simplicity and effectiveness. The essence of CF lies in capturing user preferences by learning from observed interactions. With the advent of graph neural networks (GNN) \cite{gao_gnn2-survey_RS_2023}, a variety of graph-based CF methods have emerged. For instance, NGCF \cite{wang_NGCF_SIGIR_2019} models the user-item bipartite graph, aggregating neighborhood information into embeddings via graph convolution. Building on this, GMCF \cite{su_GMCF_SIGIR_2021} proposes a graph-matching-based neural CF method, effectively capturing two types of interaction information. LightGCN \cite{he_LightGCN_SIGIR_2020} further improves performance by simplifying GCN process, retaining only neighborhood aggregation while removing feature transformation and non-linear activation. 

\medskip

\noindent\textbf{Contrastive Learning in CF}. 
Contrastive learning (CL) \cite{jing_CL-survey_2023,xu_CL_ICLR_2021} has been extensively studied in CF \cite{gao_gnn2-survey_RS_2023}, with the predominant approach relying on data augmentation. Data augmentation \cite{Bayer_Data-Augmentation-survey_CS_2022,rebuffi_Data-augmentation-survey_NeurIPS_2021} generally falls into two categories: the first is graph augmentation, while the second is embedding augmentation. In the first category, SGL \cite{wu_SGL_SIGIR_2021} generates subgraphs by dropout edges, nodes, and random walks on the interaction graph. In the second category, SimGCL \cite{yu_SimGCL_SIGIR_2022} and BIGCF \cite{zhang_BIGCF_SIGIR_2024} adds noise during the GCN process to create contrastive views. Additionally, methods like NCL \cite{Lin_NCL_WWWW_2022} and CGCL \cite{he_CGCL_SIGIR_2023} attempt to derive effective self-supervised signals through cross-layer contrast. While these methods have been proven effective, their exploration of CL loss has been limited to application. Recent works have begun to investigate the objectives of CL, such as in DirectAU \cite{wang_DirectAU_KDD_2022}, RecDCL \cite{Dan_RecDCL_WWW_2024}, and SCCF \cite{wu_SCCF_KDD_2024}. 
Although SCCF discusses the relationship between representation learning, CL, and GCN, there remains a paucity of research specifically focused on CL's core function, InfoNCE. The introduction of LightCCF addresses this gap, proposing a NA loss to enhance CF methods in aggregating neighborhood information for personalized recommendation.

\section{Conclusion} 
In this paper, we proved the neighborhood aggregation capability of the contrastive learning (CL) objective through both theoretical derivation and empirical evidence. Building on this, we introduced Light Contrastive Collaborative Filtering (LightCCF), a novel recommendation method that introduces a neighborhood aggregation objective. This objective effectively minimizes the distance between user and their interacted items while keeping user away from other positive pairs, thus achieving comprehensive neighborhood aggregation. Our experimental results reveal that LightCCF significantly outperforms existing methods on three widely used datasets. 



\begin{acks}
\vspace{0.5mm}
This work is supported by the National Natural Science Foundation of China (No.62272001) and the Open Project of Key Laboratory of Intelligent Computing \& Signal Processing of Ministry of Education (No.2023A008).

\end{acks}

\newpage

\bibliographystyle{ACM-Reference-Format}
\bibliography{sample-base}










\end{document}